# LogLog-Beta and More: A New Algorithm for Cardinality Estimation Based on LogLog Counting

Jason Qin, Denys Kim, Yumei Tung
The AOLP Core Data Service, AOL, 22000 AOL Way
Dulles, VA 20163
E-mail: jason.qin@teamaol.com

*Abstract*—The information presented in this paper defines LogLog-Beta (LogLog-β). LogLog-β is a new algorithm for estimating cardinalities based on LogLog counting. The new algorithm uses only one formula and needs no additional bias corrections for the entire range of cardinalities, therefore, it is more efficient and simpler to implement. Our simulations show that the accuracy provided by the new algorithm is as good as or better than the accuracy provided by either of HyperLogLog or HyperLogLog++. In addition to LogLog-β we also provide another one-formula estimator for cardinalities based on order statistics, a modification of an algorithm described in Lumbroso's work.

## I. Introduction

Cardinality estimation of a multiset has numerous applications in data management, data analysis, and data services, see papers [2][3][7][11][14][15], etc. Over the years, many cardinality estimation algorithms have been proposed and studied, especially the ones related to probabilistic counting [1][4][5][6][8][9][11][10][13][16][17]. In this paper we present a new algorithm related to a particular probabilistic counting family called LogLog Counting. LogLog Counting has been long studied: see the earlier ideas in Alon and Matias and Szegedy[1] and Flajolet[8], the basic LogLog and SuperLogLog algorithms in Duran and Flajolet[6], the popular HyperLogLog in Flajolet et al. [9], and the recent improvement HyperLogLog++ in Heule et al. [11].

The HyperLoglog algorithm in Flajolet et al. [9] serves as the foundation for our study. Our new algorithm, LogLog-β, can be regarded as a modification of the HyperLogLog. In this introduction we will briefly present the HyperLogLog algorithm. The remainder of the paper is organized as follows: section II details the new algorithm LogLog-β, section III gives some accuracy tests on LogLog-β, and section IV provides another one-formula only cardinality estimator and some further discussions on the two new algorithms.

HyperLogLog algorithm in Flajolet et al. [9] has five major components: data randomization by a hash function, stochastic averaging and register vector generation, the raw estimation formula, the Linear Counting in Whang and Vander-Zanden and Taylor [16], and bias corrections.

Let $S$ be a multiset of data, $M$ be a vector of size $m$ with $m = 2^p$ with $p = 4, 5, \ldots$ , $h$ be a hash function where the hash value is a $32$-bit array, let $\rho$ be a function that maps a bit array to its number of leading zeros plus one, and $\alpha_m$ be a parameter dependent of $m$ (see [9]).

HyperLogLog Algorithm:

1) Initialize $M$:

$$M[i] = 0 \text{ for } i = 0, 1, \ldots, m-1$$

2) Generate $M$:

For each $s \in S$, let $i$ be the integer formed by the left $p$ bits of $h(s)$ and $w$ be the right $32-p$ bits of $h(s)$, set

$$M[i] = max(M[i], \rho(w))$$

3) Apply the raw formula:

$$E = \frac{\alpha_m m^2}{\sum_{i=0}^{m-1} 2^{-M[i]}} \qquad (1)$$

4) Apply Linear Counting if $E \leq \frac{5}{2}m$:

$$E = m log \frac{m}{z}$$

where $z$ is the number of elements of $M$ with value $0$.

5) Apply correction if $E > \frac{1}{30}2^{32}$:

$$E = -2^{32} log\left(1 - \frac{E}{2^{32}}\right)$$

When using a 64-bit hash function, step 5 in above algorithm is no longer needed as long as the cardinality does not near $2^{64}$. In this paper we use 64-bit hash function in algorithms and simulations.

The accuracy of cardinality estimation provided by formula (1) gets better when cardinality becomes larger (proved by Flajolet et al. [9]), but the accuracy from Linear Counting degenerates as cardinality increases. Therefore, loss of accuracy around switching point $\frac{5m}{2}$, defined in [9], is inevitable. A major improvement in HyperLogLog++[11] (in addition to using 64-bit hash and sparse representation) is provided a bias correction table for a range of cardinalities to boost the accuracy around the switching point. The bias correction table and correction range are $m$ dependent and determined empirically based on the estimates offered by formula (1). In this paper HyperLogLog++ refers to the implementation without sparse representation.

The new algorithm, LogLog-β, uses only one formula and needs neither bias corrections nor Linear Counting, and therefore implementation is simplified. Our simulations show the accuracy provided by the new algorithm is as good as or better than the accuracy provided by either of HyperLogLog or HyperLogLog++ (without sparse representation).

## II. LogLog-β algorithm

In [9] Flajolet et al. proved asymptotically that the raw formula (1) of HyperLogLog has a very low standard error ( $\frac{1.04}{\sqrt{m}}$ ). However the formula fails to provide good cardinality estimation for small or pre-asymptotic cardinalities. The main idea of the new algorithm, LogLog-β, is to modify the raw formula and to make it applicable to the entire range of cardinalities, from very small to very large.

### A. LogLog-β algorithm

LogLog-β Algorithm:

1) Initialize $M$:

Set $M[i] = 0$ for $i = 0, 1, \ldots, m-1$

2) Generate $M$:

For each $s \in S$, let $i$ be the integer formed by the left $p$ bits of $h(s)$ and $w$ be the right 64-p bits of $h(s)$, set

$$M[i] = max\ (M[i], \rho(w)\ )$$

3) Apply the formula:

$$E = \frac{\alpha_m m(m-z)}{\beta(m,z) + \sum_{i=0}^{m-1} 2^{-M[i]}} \qquad (2)$$

where $z$ is the number of elements of $M$ with value $0$ and $\beta(m,z)$ is a function of $m$ and $z$.

Clearly, formula (2) is a modification of formula (1) which replaces the numerator $\alpha_m m^2$ with $\alpha_m m(m-z)$ and adds a term $\beta(m,z)$ to the denominator. Furthermore, formula (2) and (1) are identical when $z = 0$ and $\beta(m,0) = 0$.

### B. Function $\beta(m,z)$

The function $\beta(m,z)$ is a kind of bias minimizer and should go to $0$ when $z$ goes to 0. Therefore, for very large cardinalities, formula (2) is almost the same as formula (1). The ultimate goal is to find $\beta(m,z)$ such that formula (2) could yield highly accurate estimations for the entire range of cardinalities.

There are many ways to choose $\beta(m, z)$, for example $\beta(m,z) = \beta_0(m)z$, or $\beta(m,z) = \beta_0(m)z + \beta_1(m)z^2 + \ldots$. For the sake of convenience of discussion in this paper we limit $\beta(m,z)$ to the following form:

$$\beta(m,z) = \beta_0(m)z + \beta_1(m)z_l + \beta_2(m){z_l}^2 + \ldots + \beta_k(m){z_l}^k$$

where $z_l = log\ (z+1)$, $k \geq 0$, and $\beta_0(m)$, $\beta_1(m), \ldots, \beta_k(m)$ are $m$ dependent constants. Clearly $\beta(m,z) = 0$ when $z = 0$. An implementation, based on Horner's rule, $\beta(m,z)$ could be evaluated by a total of $(k+1)$ multiplications and $k$ additions when $z_l$ is provided.

### C. Constants $\beta_0(m)$, $\beta_1(m), \ldots, \beta_k(m)$

For given $m$, $k$, and a data set of cardinality $c$, under the best circumstances we expect $\beta(m,z) = \hat{\beta}(m,z)$ where

$$\hat{\beta}(m,z) = \frac{\alpha_m m(m-z)}{c} - \sum_{i=0}^{m-1} 2^{-M[i]}$$

If we randomly generate data sets with given cardinalities $c_1$, $c_2$, …, and $c_n$ (from very small to very large) and compute $z$ and $\hat{\beta}(m,z)$ for each $c_i$, then by solving a least square problem $min\ \|\beta(m,z) - \hat{\beta}(m,z)\|_2^2$ we shall be able to determine $\beta_0(m)$, $\beta_1(m), \ldots, \beta_k(m)$. In practice we pick cardinalities $c_i$ such that $c_1 < c_2 < \ldots < c_n$, equally distanced, with $n \gg k$ and $z = 0$ for some of the large cardinalities. Furthermore for each given $c_i$, we compute the means of $z$ and $\hat{\beta}(m, z)$ over many randomly generated data sets, then use the means to solve the least square problem.

In our study we used MURMUR3-64 as the hash function and java.util.Random for generating data sets with given cardinalities. For example for $p = 14$, $m = 2^{14}$, we computed a few $\beta(m, z)$:

$$\beta(m,z) = -0.309142z + 13.733192z_l - 8.636985z_l^2 + 1.328973z_l^3$$

$$\beta(m,z) = -0.350308z + 3.949176z_l - 6.403082z_l^2 + 3.289908z_l^3 - 0.643495z_l^4 + 0.051568z_l^5$$

The number of terms of $\beta(m, z)$ should be used is determined by accuracy requirement: the larger $k$ the better accuracy the algorithm provides. However, we cannot reach arbitrary accuracy by simply increasing $k$: the optimal accuracy that can be achieved is dictated by $m$, the size of vector $M$. In our study we found 3 to 7 to be a reasonable range for $k$.

## III. Accuracy tests of LogLog-β

As long as $\beta(m, z)$ is properly determined, the LogLog-β algorithm can provide estimation for different ranges of cardinalities with accuracy comparable to HyperLogLog++. Namely, formula (2) can perform as well as Linear Counting for small cardinalities and HyperLogLog Raw with bias correction for large and very large cardinalities.

For completeness, we should determine $\beta(m, z)$ and perform accuracy tests for each precision $p$ ($m = 2^p$). However, since the processes are quite similar, this paper only shows the tests for precision $p = 14$ ($m = 2^{14} = 16384$). For very large cardinalities the estimations by HyperLogLog, LogLog-β, and HyperLogLog++ are literally the same therefore cardinalities are limited to $\leq 200,000$ in accuracy tests.

To determine $\beta(m, z)$ for $p = 14$, we computed the means of $z$ and $\hat{\beta}(m,z)$ over 100 (the more the better) randomly generated data sets per cardinality for cardinalities from $1,000$ to $170,000$ in increments of $1,000$. With $k = 7$ (8 terms) the function $\beta(2^{14},z)$ is

$$-0.370393914z + 0.070471823z_l + 0.17393686z_l^2 + 0.16339839z_l^3 - 0.09237745z_l^4 + 0.03738027z_l^5 - 0.005384159z_l^6 + 0.00042419z_l^7$$

and it is used for $p = 14$ in all the accuracy tests in this study.

By using above $\beta(m, z)$ and any of the hash functions MURMUR3, MD5, and SHA, we found the test results of LogLog-β to be almost the same. Specifically, the function $\beta(m, z)$ was calculated using one hash function could be used in LogLog-β in place of any other hash function as long as their hash values are uniformly distributed and well randomized. We chose MMURMUR3 for this study because of its performance.

Both tests on the mean of relative errors and the mean of absolute values of relative errors for 500 randomly generated datasets per cardinality show that LogLog-β provides slightly better estimations than HyperLogLog and HyperLogLog++, especially for the cardinalities in the range of 10,000 to 85,000, see Fig. 1 and Fig. 2. Please note that LogLog-β has better performance in terms of accuracy and stability over Linear Counting for almost all the small to mid-range cardinalities. Figure 3 shows the relative error of one randomly generated dataset per cardinality.

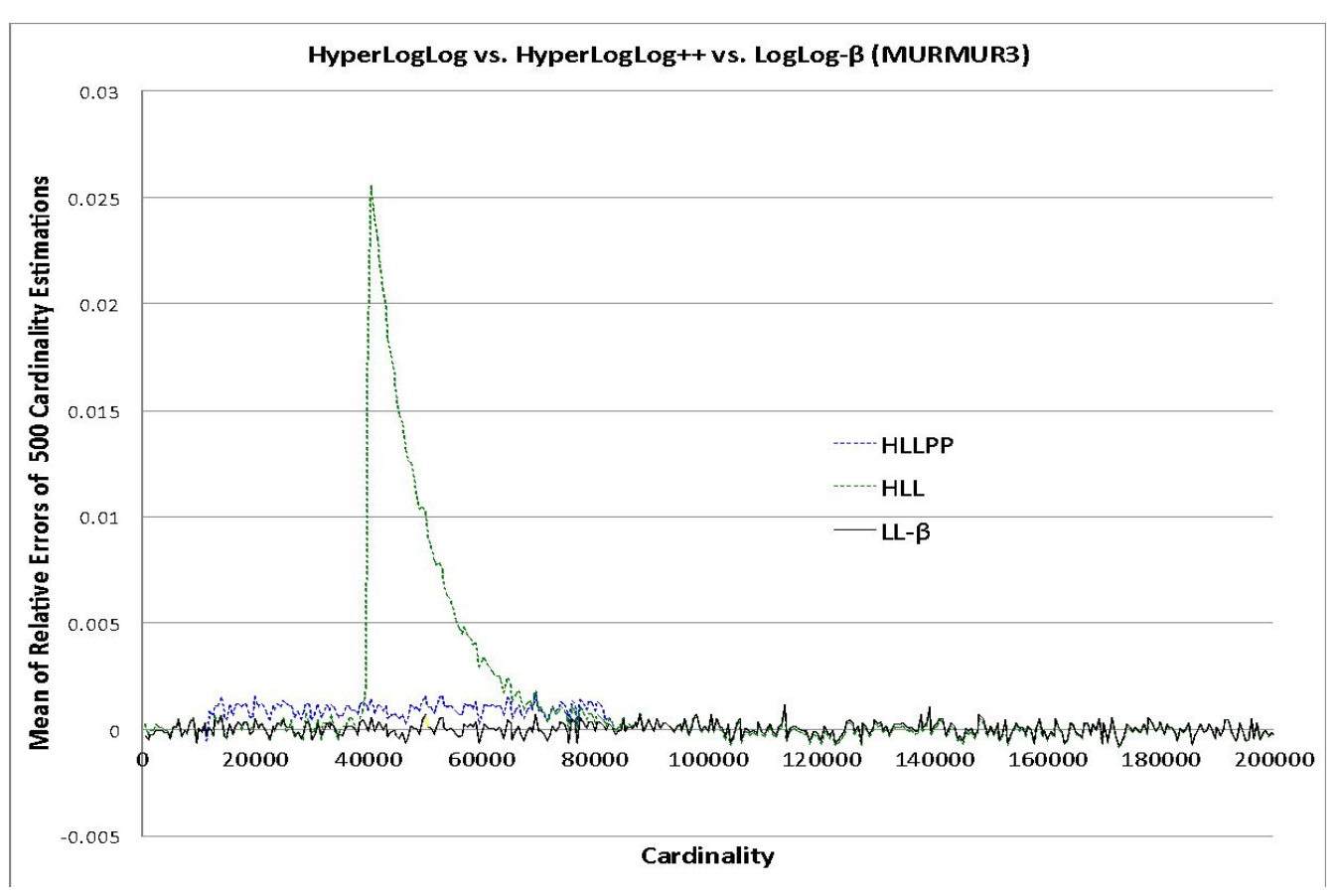

Fig.1. The mean of relative errors of cardinality estimations for 500 randomly generated datasets per cardinality (the $x$-axis). Tested cardinalities are from 500 to 200,000 in every 500.

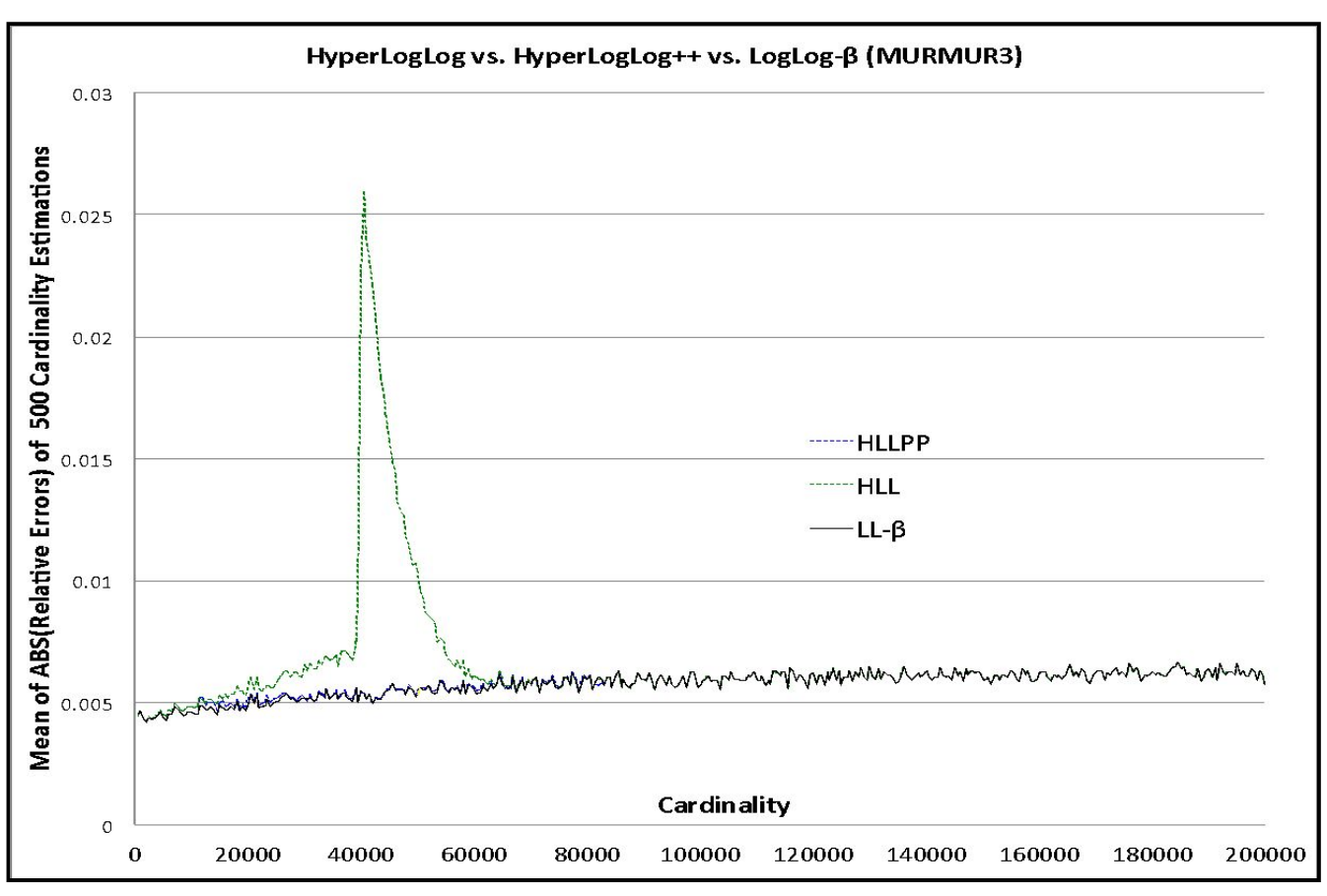

Fig. 2. The mean of abs(relative errors) of cardinality estimations for 500 randomly generated datasets per cardinality (the $x$-axis). Tested cardinalities are from 500 to 200,000 in every 500.

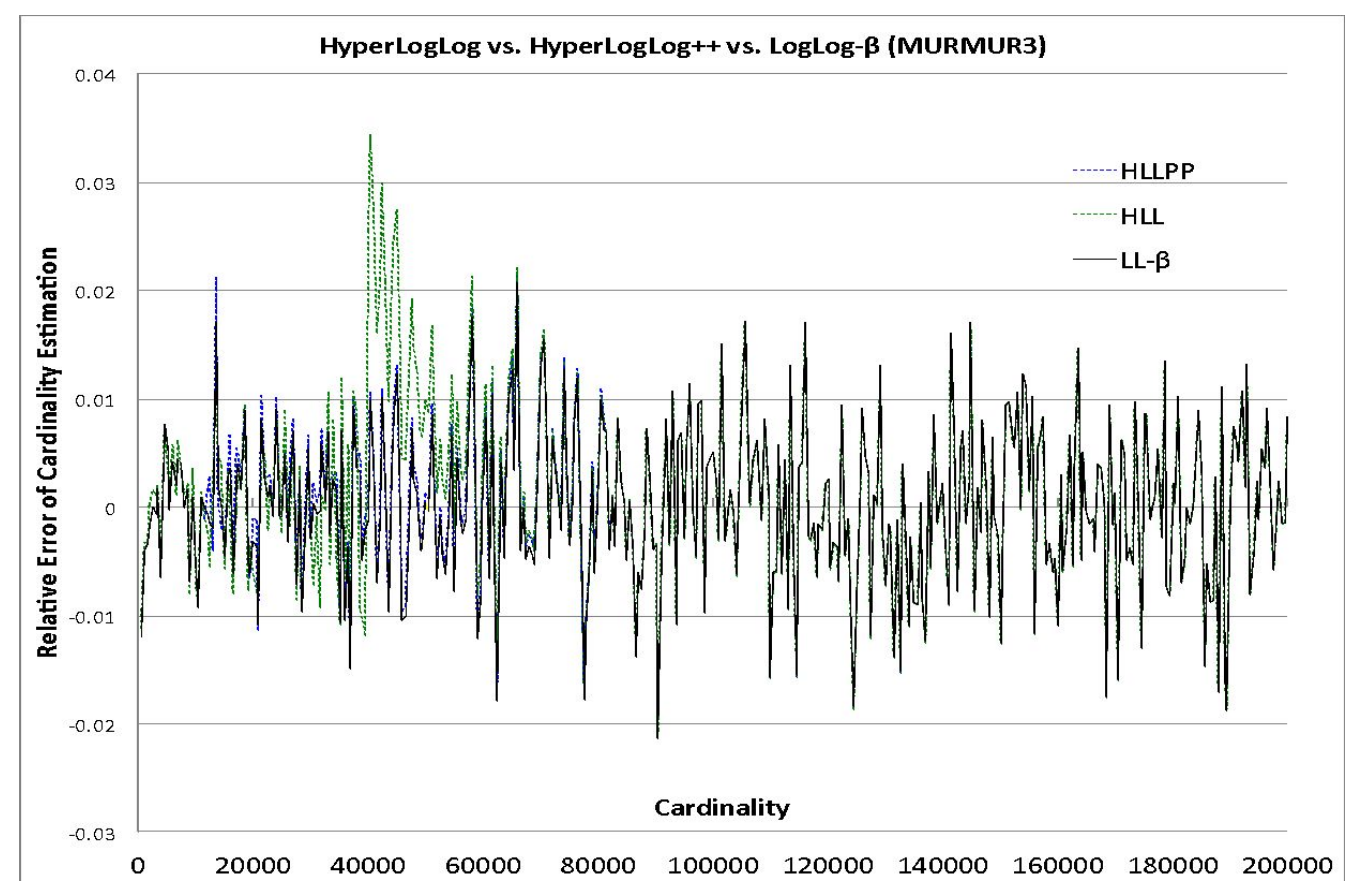

Fig. 3. The relative error of the cardinality estimation of one randomly generated dataset per cardinality (the $x$-axis). Tested cardinalities are from 500 to 200,000 in every 500.

Figure 4, 5, and 6 show the empirical histograms of cardinality estimations for 500 randomly generated datasets per cardinality with cardinality being 1,000; 50,000; and 100,000 respectively. Both HyperLogLog and HyperLogLog++ use the same formulas in Fig. 4 and 6: Linear Counting for cardinality = 1,000 and formula (1) (with a minor bias correction for HyperLogLog++) for cardinality= 100,000. Therefore, the histograms corresponding to HyperLogLog and HyperLogLog++ are almost identical in Fig. 4 and 6. In both figures LogLog-β shows comparable or slightly better behaviors. In Figure 5 HyperLogLog, HyperLogLog++, and LogLog-β show different behaviors since HyperLogLog uses formula (1) and HyperLogLog++ uses both formula (1) and bias correction.

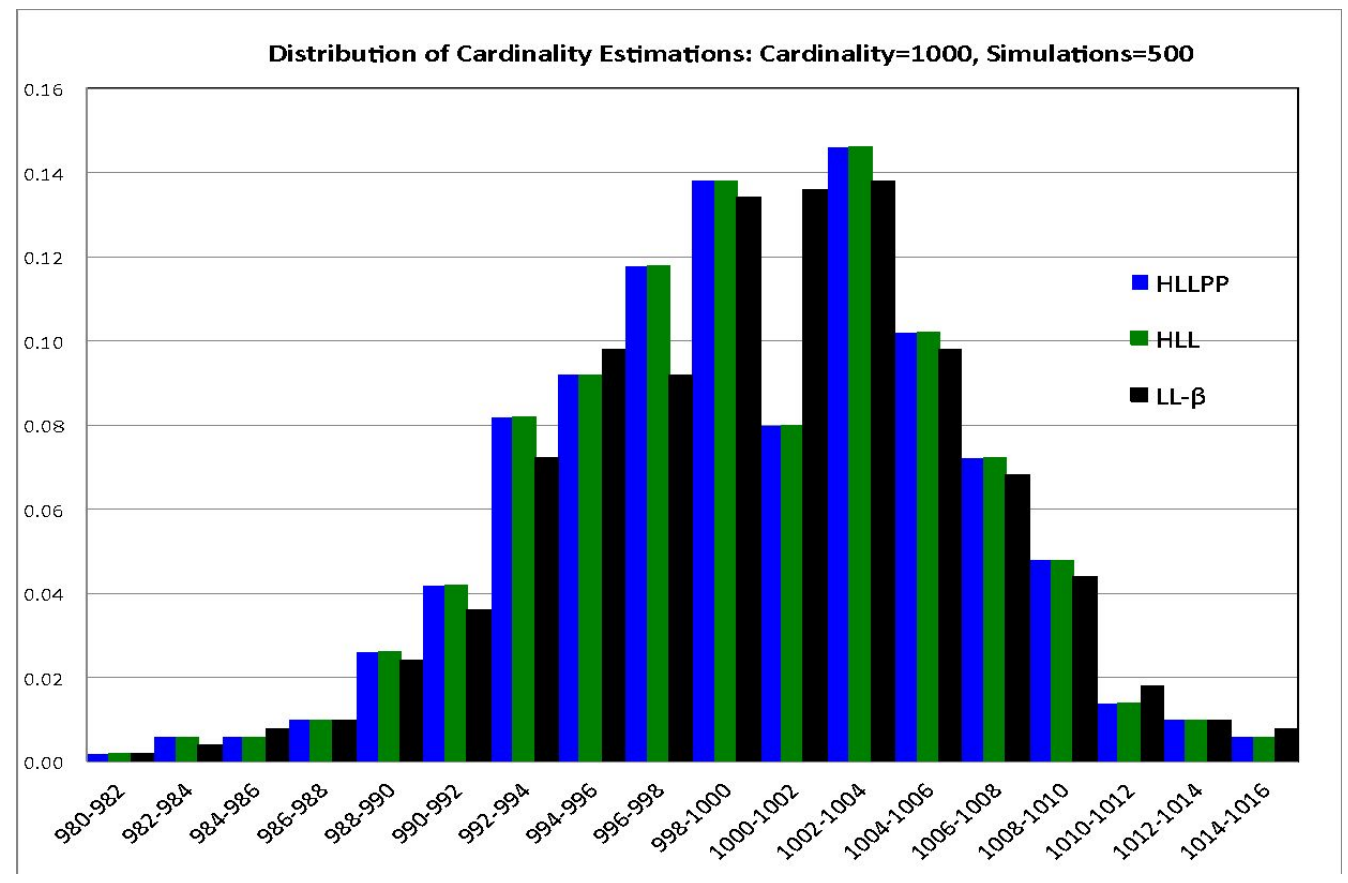

Fig. 4. The histogram of the cardinality estimations of 500 randomly generated datasets for cardinality 1,000.

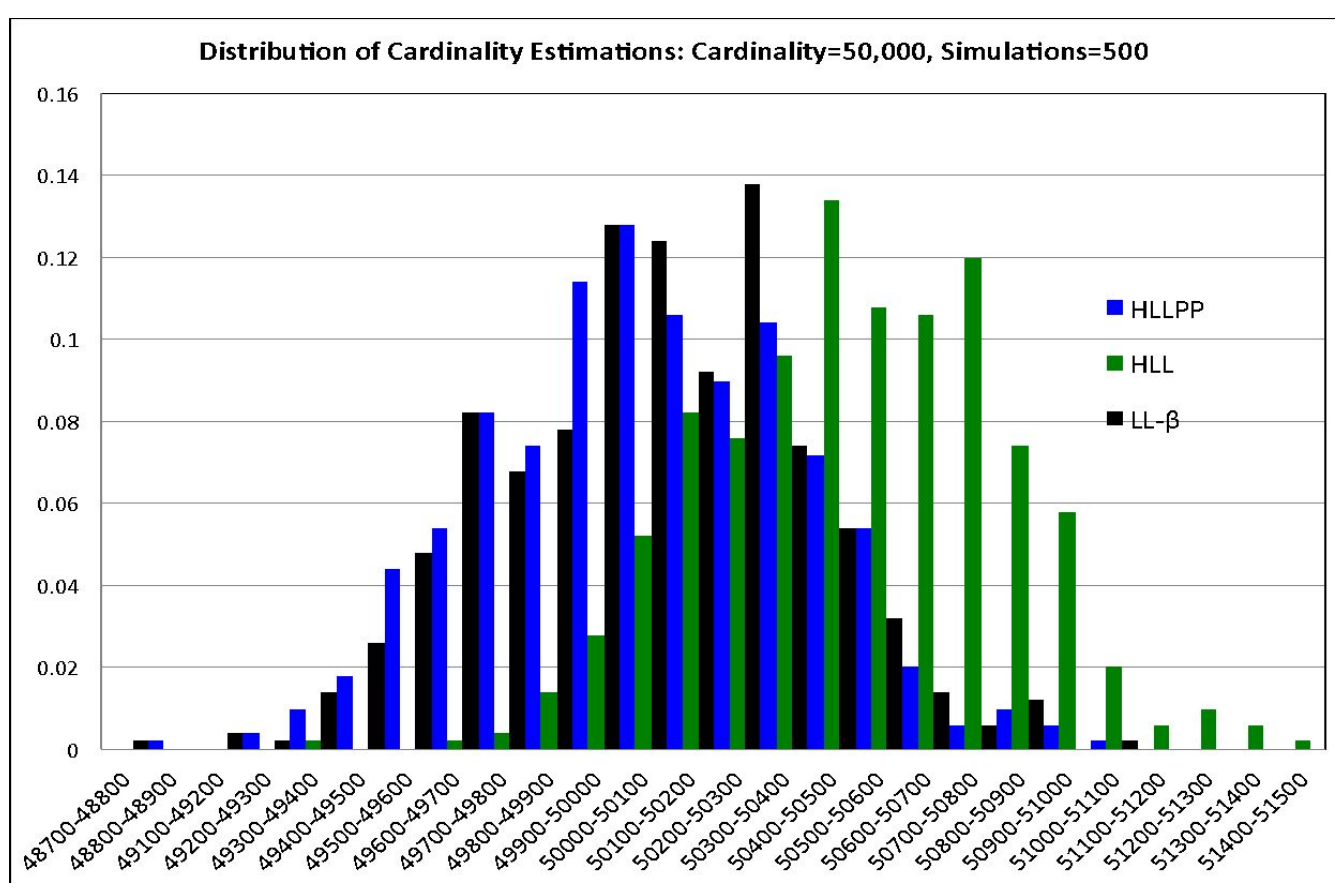

Fig. 5. The histogram of the cardinality estimations of $500$ randomly generated datasets for cardinality $50,000$.

## IV. More discussions

The formula (2) of algorithm LogLog-β requires two major changes compared to HyperLogLog Raw (1), and these two changes make the formula capable of covering the entire range of cardinalities with comparable or better accuracy. Replacing $m^2$ with $m(m-z)$, makes no difference for the very large cardinalities but has significant impact on the estimates of small and mid-range cardinalities. It forces the estimates to roughly align with the real cardinalities. The other change, adding a term $\beta(m, z)$ to the denominator, further corrects the remaining error/bias. The combination of the two changes makes formula (2) work more accurately for the entire range of cardinalities. Both changes are essential, and the second change could have many variations and forms.

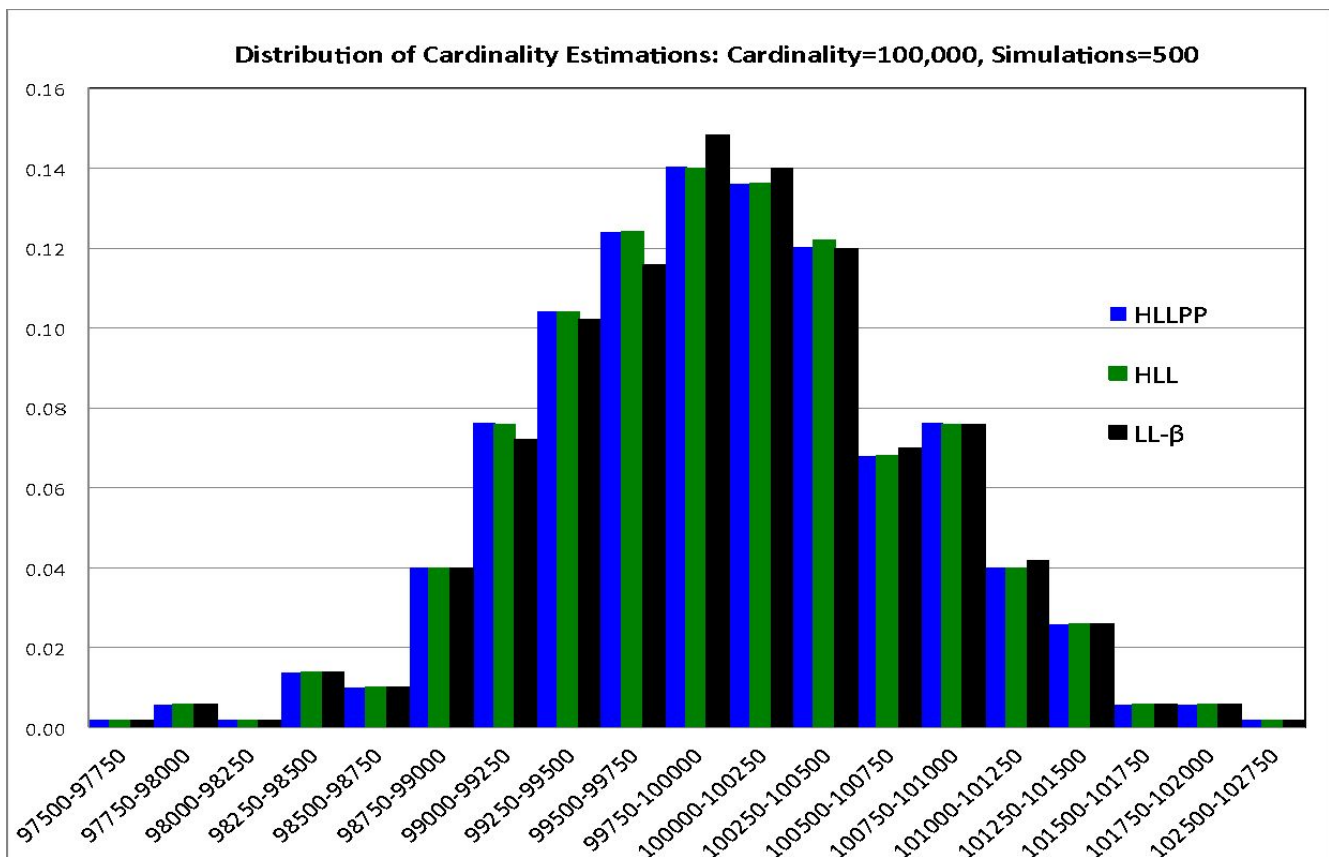

Fig. 6. The histogram of the cardinality estimations of $500$ randomly generated datasets for cardinality $100,000$.

In terms of implementation and computation complexity, LogLog-β offers more advantages: one formula for the entire range of cardinalities, a much simpler process flow, and no bias correction and lookup tables. Furthermore, for any existing implementations of HyperLogLog or HyperLogLog++ (without sparse representation), LogLog-β requires no regeneration of vectors $M$ and it can be applied directly to the existing vectors. Therefore, switching to LogLog-β requires a very small effort and produces comparable or better results.

The idea of replacing $m^2$ by $m(m-z)$ can be applied to other cardinality estimation algorithms as well. For example, we apply this idea to an unbiased optimal cardinality estimation algorithm proposed and studied in Lumbroso[13]. The algorithm, like HyperLogLog, performs well for very large cardinalities but depends on Linear Counting and bias corrections for small and pre-asymptotical cardinalities. The core algorithm in Lumbroso[13], without Linear Counting and bias correction, is described as follows:

Let $S$, $M$, $m$, and $p$ be the same as described in HyperLogLog algorithm, but $h$ be a hash function with hash value in interval $(0, 1)$,

Lumbroso's Core Algorithm:
1) Initialize $M$:
Set $M[i] = 1$ for $i = 1, \ldots, m$
2) Generate $M$:
For each $s \in S$, set $y = h(s)$, $i = \lfloor ym \rfloor + 1$, and
$M[i] = min\,(M[i], ym - \lfloor ym \rfloor)$
3) Apply the core formula:

$$E = \frac{m(m-1)}{\sum_{i=1}^{m} M[i]} \tag{3}$$

It is proved by Lumbroso[13] that the above algorithm yields optimal estimation for very large cardinalities. To make it work for small or mid-range cardinalities we substitute $1$ in formula (3) with $z$ which results in a new formula:

$$E = \frac{m(m-z)}{\sum_{i=1}^{m} M[i]} \tag{4}$$

where $z$ is the number of $M[i]$ with value 1. There is no need to add a correction term to the denominator in this case. The estimations provided by formula (4) are strikingly accurate for all ranges of cardinalities, especially for the small and mid-range cardinalities. In Figure 7 we show some test results on this new algorithm, called MMV (Mean of Minimum Values).

Formula (2) of LogLog-β and formula (4) of MMV, derived from different approaches (counting vs. ordering), are deeply related (detailed study is in a separate paper). The term $2^{-M[i]}$ of (2) could be regarded as an approximation of the term $M[i+1]$ of (4). For a given $m$ both algorithms provide comparable accuracies, but LogLog-β requires less memory.

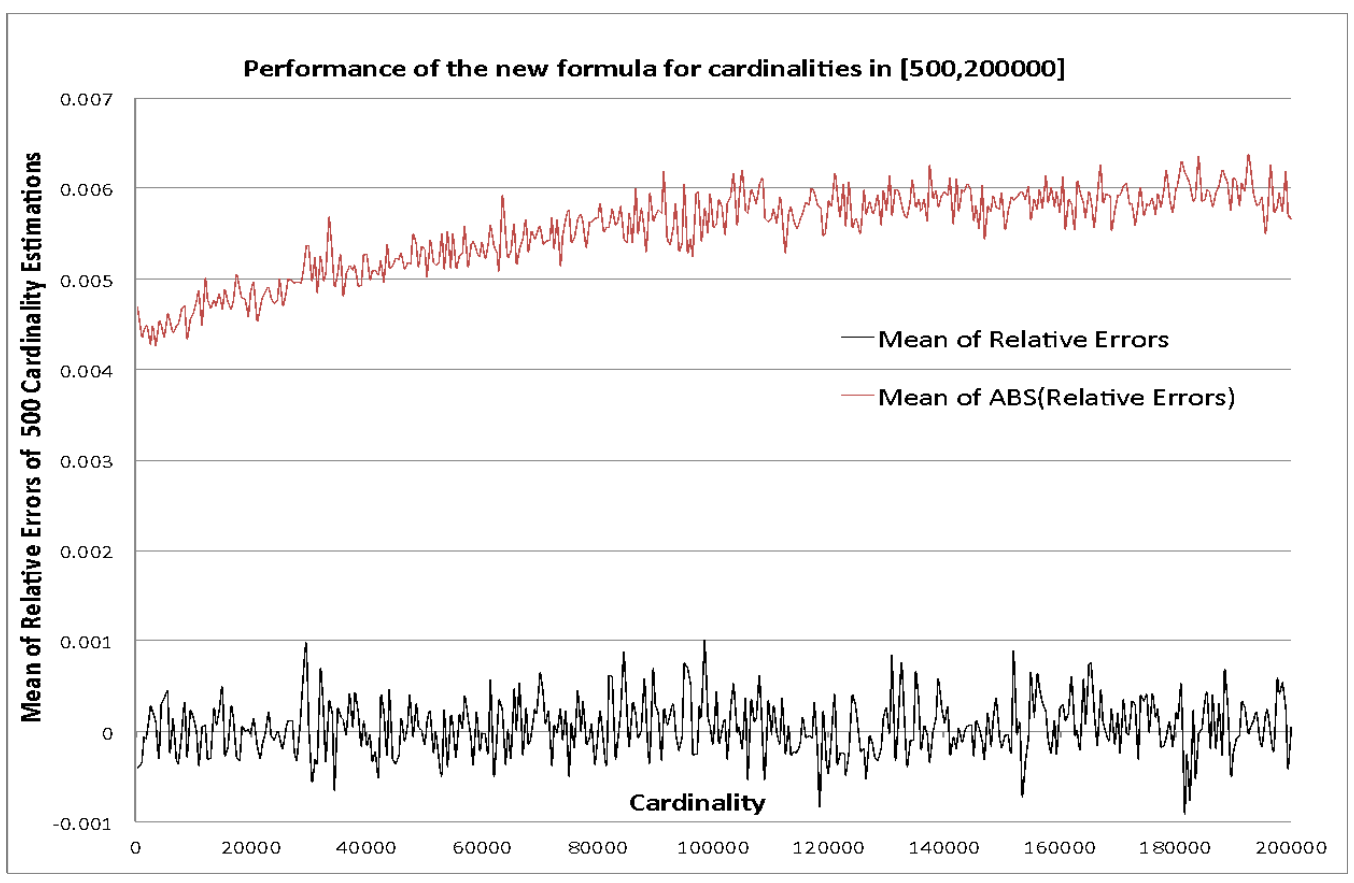

Fig. 7. Cardinality estimations of $500$ randomly generated datasets for each cardinality in every $500$ from $500$ to $200,000$.

## Appendix: Coefficients of $\beta(m,z)$

In the paper we only listed the coefficients of $\beta(m,z)$ for $p = 14$ and $k = 7$ , however over the years people were interested in $\beta(m,z)$ for other $p$ values too. We provide here some of the coefficients of $\beta(m,z)$ calculated while the paper was written in 2016. As suggested in the paper that different hash function, random number generator, and cardinality sampling points could yield different coefficients.

$[\beta_0, \beta_1, \beta_2, ...\beta_7]$ for $m = 2^p$ , $p = 4, 5, 6, ..., 18$ and $k = 7$ are defined as:

p=4

```
[-0.582581413904517,-1.935300357560050,11.07932375
8035073,-22.131357446444323,22.505391846630037,-12
.000723834917984,3.220579408194167,-0.342225302271
235]
```

p=5

```
[-0.7518999460733967,-0.9590030077748760,5.5997371
322141607,-8.2097636999765520,6.5091254894472037,-
2.6830293734323729,0.5612891113138221,-0.046333162
2196545]
```

p=6

```
[29.8257900969619634,-31.3287083337725925,-10.5942
523036582283,-11.5720125689099618,3.81887543739074
92,-2.4160130328530811,0.4542208940970826,-0.05751
55452020420]
```

p=7

```
[2.8102921290820060,-3.9780498518175995,1.31626800
41351582,-3.9252486335805901,2.0080835753946471,-0
.7527151937556955,0.1265569894242751,-0.0109946438
726240]
```

p=8

```
[1.00633544887550519,-2.00580666405112407,1.643697
49366514117,-2.70560809940566172,1.392099802442225
98,-0.464703742721831 90,0.07384282377269775,-0.005
78554885254223]
```

p=9

```
[-0.09415657458167959,-0.78130975924550528,1.71514
946750712460,-1.73711250406516338,0.86441508489048
924,-0.23819027465047218,0.03343448400269076,-0.00
207858528178157]
```

p=10

```
[-0.25935400670790054,-0.52598301999805808,1.48933
034925876839,-1.29642714084993571,0.62284756217221
615,-0.15672326770251041,0.02054415903878563,-0.00
112488483925502]
```

p=11

```
[-4.32325553856025e-01,-1.08450736399632e-01,6.091
56550741120e-01,-1.65687801845180e-02,-7.958293410
87617e-02,4.71830602102918e-02,-7.81372902346934e-
03,5.84268708489995e-04]
```

p=12

```
[-3.84979202588598e-01,1.83162233114364e-01,1.3039
6688841854e-01,7.04838927629266e-02,-8.95893971464
453e-03,1.13010036741605e-02,-1.94285569591290e-03
,2.25435774024964e-04]
```

p=13

```
[-0.41655270946462997,-0.22146677040685156,0.38862
131236999947,0.45340979746062371,-0.36264738324476
375,0.12304650053558529,-0.01701540384555510,0.001
02750367080838]
```

p=14

```
[-3.71009760230692e-01,9.78811941207509e-03,1.8579
6293324165e-01,2.03015527328432e-01,-1.16710521803
686e-01,4.31106699492820e-02,-5.99583540511831e-03
,4.49704299509437e-04]
```

p=15

```
[-0.38215145543875273,-0.89069400536090837,0.37602
335774678869,0.99335977440682377,-0.65577441638318
956,0.18332342129703610,-0.02241529633062872,0.001
21399789330194]
```

p=16

```
[-0.37331876643753059,-1.41704077448122989,0.40729
184796612533,1.56152033906584164,-0.99242233534286
128,0.26064681399483092,-0.03053811369682807,0.001
55770210179105]
```

p=17

```
[-0.36775502299404605,0.53831422351377967,0.769702
89278767923,0.55002583586450560,-0.745755882611469
41,0.25711835785821952,-0.03437902606864149,0.0018
5949146371616]
```

p=18

```
[-0.36479623325960542,0.99730412328635032,1.553543
86230081221,1.25932677198028919,-1.533259482091101
63,0.47801042200056593,-0.05951025172951174,0.0029
1076804642205]
```